\documentclass[12pt]{article}

\usepackage[authoryear]{natbib}
\bibpunct{(}{)}{,}{a}{,}{;} 

\usepackage[english]{babel}
\usepackage[utf8]{inputenc}
\usepackage[T1]{fontenc}

\usepackage{float}
\usepackage{multirow,multicol}

\usepackage{bm}
\usepackage{amsmath}
\usepackage{amsfonts}
\usepackage{amsthm} 

\usepackage[ruled,vlined,resetcount]{algorithm2e}
\SetKw{KwFrom}{from}
\SetKwHangingKw{Point}{\raisebox{0.25ex}{\scalebox{0.5}{$\blacksquare$}}}
\SetKwInOut{Param}{Parameters}
\SetKw{Return}{Return}

\usepackage{mathrsfs} 

\usepackage{mdframed} 
\newmdenv{allfour}
\newmdenv[leftline=false,rightline=false]{topbot}
\newmdenv[topline=false,rightline=false]{leftbot}
\newmdenv[topline=false,leftline=false]{rightbot}
\newmdenv[topline=false,rightline=false,leftline=false]{bottom}

\usepackage[justification=centering]{caption}
\usepackage{subcaption} 

\usepackage[pdftex]{graphicx}
\DeclareGraphicsExtensions{.jpg,.pdf,.png}

\usepackage[pdftex,colorlinks=true,linkcolor=blue,citecolor=blue,urlcolor=blue]{hyperref}
\usepackage{cleveref}

\usepackage{anysize}
\marginsize{25mm}{25mm}{15mm}{25mm}

\usepackage{icomma}

\usepackage{xcolor, colortbl}

\usepackage{authblk}

\usepackage[title]{appendix}

\newcommand{\q}[1]{``#1''}%
\newcommand{\peq}{\; .} 
\newcommand{\veq}{\; ,} 
\newcommand{\R}{\mathbb{R}} 

\newcommand{\e}{\mathbb{E}} 
\newcommand{\cov}{\textrm{Cov}} 
\newcommand{\var}{\textrm{Var}} 

\newcommand{\spn}[1]{\textrm{span}\left\lbrace #1\right\rbrace}%

\newtheorem{proposition}{Proposition}[section]

\providecommand{\keywords}[1]
{
  \small    
  \textbf{\textit{Keywords --}} #1
}

\begin{document}

\title{A matrix-free approach to geostatistical filtering}

\author[1]{Mike PEREIRA%
  \thanks{Contact: \texttt{mikep@chalmers.se}}}  
\author[2]{Nicolas DESASSIS}
\author[3]{C\'edric MAGNERON}
\author[4]{Nathan PALMER}

\affil[1]{Chalmers University of Technology and University of Gothenburg \\
Gothenburg, Sweden}
\affil[2]{MINES ParisTech, PSL Research University\\
   Fontainebleau, France}
\affil[3]{Estimages\\
   Paris, France}
\affil[4]{Central Petroleum Ltd\\
          Brisbane, Australia}


\date{February 25th, 2020}

\maketitle

\begin{abstract}
In this paper, we present a novel approach to geostatistical filtering which tackles two challenges encountered when applying this method to complex spatial datasets: modeling the non-stationarity of the data while still being able to work with large datasets. The approach is based on a finite element approximation of Gaussian random fields expressed as an expansion of the eigenfunctions of a Laplace--Beltrami operator defined to account for local anisotropies. The numerical approximation of the resulting random fields using a finite element approach is then leveraged to solve the scalability issue through a matrix-free approach. Finally, two cases of application of this approach, on simulated and real seismic data are presented. \\
\end{abstract}


\section{Introduction}

When trying to characterize an oil reservoir in the subsurface of a field, a go-to method consists in using seismic imaging. Through this process, an echography of the subsurface is obtained by \q{shooting} acoustic waves into the subsurface, and then studying their reflection on the different geological interfaces composing the subsurface. In particular, complex processing methods are performed to turn the detected times of arrival of these waves after reflection on a given interface into the actual depth of this surface. The methods applied to acquire the seismic data often add noise to the data. If not removed, it is very challenging to identify zones of interest in the subsurface. Like the signal, the noise tends to be spatially correlated inside the \q{seismic} cube. In this context, Geostatistics provide tools and methods suited to model such spatially structured data.

Indeed, Geostatistics is the branch of Statistics that focuses on variables defined over a spatial domain \citep{chiles1999geost}. Such variables are usually observed on a set of fixed locations in the domain and are spatially structured. For instance, for seismic data, the variable is the measured amplitude and the spatial domain is formed by the domain of acquisition and the seismic depth. When referring to spatial structure, we mean here that the similarity of observations of the variable is linked to their locations within the spatial domain. Such variables are modeled as (realizations of) spatially correlated (Gaussian) random fields. This probabilistic model for the variable then allows, once a dependency between correlation and space is fixed, to mimic the spatial structure of the data, to perform tasks such as the simulation of variables with the same spatial structure, estimating the variable at unobserved locations or, as illustrated in this paper, filtering out a specific spatial structure of the variable \citep{wackernagel2013multivariate}.

Geostatistical filtering is a method used in many applications dealing with spatial data affected by spatially structured noises \citep{goovaerts2004accounting,bourennane2012geostatistical}, among which seismic processing \citep{hoeber2003use,piazza2015m}. It relies on the assumption that the noisy signal at hand results from a complex phenomenon that can be seen as a superposition of independent simpler phenomena. Hence the noisy signal is written as a sum of independent signal and noise structures, each one characterized by its own range of influence and spatial structure: filtering consists in extracting the signal structure (considered as the \q{noise-free} signal) from observations of the overall sum.

Within the geostatistical framework, the components of the observed signal are modeled as independent (realizations of) Gaussian random fields defined over the spatial domain. These random fields should mimic even complex spatial structures observed in the data. For instance, when dealing with seismic data, these structures relate either to the underlying geometry of layers composing the subsurface, or to the acquisition process of the data. Hence, suitable models are needed to produce realizations with similar properties.

Besides, even if this modeling problem is taken care of, one might be confronted to scalability issues. For instance, geostatistical filtering usually requires to build matrices as big as the number of observation points and to solve the associated linear systems  \citep{wackernagel2013multivariate}. For seismic datasets, this size is that of the observation grid and therefore can easily amount to several millions of points. Hence, the filtering scheme should be scalable both memory-wise and computationally-wise if we are to apply them to such datasets.

In this paper, we present a full framework allowing to perform the geostatistical filtering of large complex datasets while tackling the two challenges described above. The outline is as follows. In \Cref{sec:theory}, we review the theory surrounding geostatistical filtering, and the main modeling and computational problems it poses. We then propose in \Cref{sec:mf} a novel approach to geostatistical filtering, based on theoretical results obtained by \citet{pereiraPhd2019} and which can be seen as an extension of the SPDE approach initially proposed by \citet{lindgren2011explicit}. We finally illustrate in \Cref{sec:appli} the use of this approach on two case studies: a synthetic one based on simulated data, and a real one based on actual seismic data.


\section{Geostatistical filtering}
\label{sec:theory}


\subsection{Basic geostatistical modeling}
\label{sec:geos_model}

\subsubsection{Isotropic stationary Gaussian random fields}

In the simplest geostatistical modeling, the variable of interest $z$, defined over a spatial domain $\mathcal{D}$, is assumed to be a particular realization of a Gaussian random field $Z$ also defined over $\mathcal{D}$. This means in particular that $Z$ can be entirely characterized through its first two moments. Assuming now that the expectation of $Z$ is constant across $\mathcal{D}$, i.e. 
\begin{equation}
\exists m\in\R, \; \forall x\in\mathcal{D}, \quad \e[Z(x)]=m \veq
\end{equation}
its spatial structure can be accurately described using a covariance function $C_Z$ which associates to each pair of points $x_1, x_2 \in \mathcal{D}$ the covariance of $Z(x_1)$ and $Z(x_2)$:
\begin{equation}
\begin{aligned}
C_Z(x_1, x_2)&= \cov\left(Z(x_1),Z(x_2)\right) \\
&=\e\left[\left(Z(x_1)-m\right)\left(Z(x_2)-m\right)\right] 
,\quad x_1, x_2 \in \mathcal{D}\peq
\end{aligned}
\end{equation}
In this paper, only zero-mean Gaussian random fields will be considered (i.e. $m=0$).

Of particular interest is the case where we further assume that for any $x_1, x_2 \in \mathcal{D}$, $C_Z$ is a radial function of the separating vector $x_2-x_1$, i.e., that there exists some function $C_0:\mathbb{R}_+ \rightarrow \mathbb{R}$ satisfying
\begin{equation}
\forall x_1, x_2 \in \mathcal{D}\subset \mathbb{R}^d, \quad
C_Z(x_1, x_2)=C_0(\Vert x_2-x_1\Vert) \veq
\label{eq:rad_cov}
\end{equation}
where $\Vert\cdot\Vert$ denotes the usual Euclidean norm on $\mathbb{R}^d$. In this case, $Z$ is said to be second-order stationary (given that its first two moments are invariant by translation) and isotropic (given that these same moments are invariant by rotation). In particular, the function $C_0$ in \eqref{eq:rad_cov} must be such that $C_Z$ still defines a covariance function. Bochner's theorem, actually provides a characterization of continuous covariance functions as the inverse Fourier transform of positive bounded symmetric measures \citep{chiles1999geost}. This theorem can in turn be used to yield a sufficient condition on $C_0$ for \eqref{eq:rad_cov} to define a \q{valid} covariance function.

\begin{proposition}[Corollary of Bochner's theorem]
If $C_0 : \R_+\rightarrow\R$ is such that the Fourier transform (on $\R^d$) of the function $\left(h \in\R^d \mapsto C_0(\Vert h \Vert)\right)$ defines a non-negative, radial and integrable function on $\R^d$ then the function $C_Z$ defined through \eqref{eq:rad_cov} is a continuous covariance function. 
\label{prop:cor_bochner}
\end{proposition}

When the conditions of \Cref{prop:cor_bochner} are satisfied, the function $f_0 : \R_+ \rightarrow \R$ defined by 
\begin{equation}
\forall \xi\in\R^d, \quad \mathcal{F}\left[ h \mapsto C_0(\Vert h \Vert)\right](\xi)=f_0(\Vert\xi\Vert) \veq
\end{equation}
is called spectral density. In particular, \citet{ORMEROD1979559} gives a formula linking a radial covariance function $C_0$ (which should be both integrable and square-integrable as a radial function of $\R^d$) and its associated spectral density $f_0$:
\begin{equation}
f_0(\Vert \xi\Vert)=\frac{1}{(2\pi)^{d/2}}\Vert \xi\Vert^{1-d/2}\int_0^{\infty}C_0(r)J_{d/2-1}(\Vert \xi\Vert r)r^{d/2}dr, \quad \xi\in\R^d \veq
\label{eq:sd_radial}
\end{equation}
where $J_{d/2-1}$ denotes the J-Bessel function with parameter $d/2-1$. Conversely, the expression of the radial covariance function $C_0$ can be retrieved from its radial spectral density $f_0$ through
\begin{equation}
C_0(\Vert h\Vert)=(2\pi)^{d/2}\Vert h\Vert^{1-d/2}\int_0^{\infty}f_0(r)J_{d/2-1}( \Vert h\Vert r)r^{d/2}dr, \quad h\in\R^d \peq
\label{eq:cov_radial}
\end{equation}

\Cref{prop:cor_bochner} can be used to derive parametrized families of isotropic covariance functions $C_0$ that are classically used in geostatistical modeling. Catalogs of such functions (an associated spectral densities) can be found for instance in the books of \citet{chiles1999geost} and \citet{lantuejoul2013geostatistical}. Note in particular that linear combinations  with positive coefficients of such functions still define appropriate covariance functions.

\subsubsection{Covariance modeling}

Recall now that we have at our disposal observations of a realization $z$ of $Z$ at some locations $x_1, \dots, x_{n}\in\mathcal{D}$. Determining a suitable combination of isotropic covariance functions to model the covariance of $Z$ from this observation can be done using variogram modeling. The semi-variogram function of $Z$ is defined as the function $\gamma_Z : \mathcal{D}\times\mathcal{D}\rightarrow \R_+$ such that
\begin{equation}
\gamma_Z(x_1,x_2)=\frac{1}{2}\var\left[Z(x_1)-Z(x_2)\right], \quad x_1,x_2\in\mathcal{D} \peq
\end{equation}  
Within the assumption that $Z$ is second-order stationary and isotropic, its semi-variogram is  once again a radial function $\gamma_0$ of the separating vector, thus giving:
\begin{equation}
\gamma_Z(x_1,x_2)=\gamma_0(\Vert x_2-x_1\Vert), \quad x_1,x_2\in\mathcal{D} \peq
\end{equation}
Also, the isotropic semi-variogram function $\gamma_0$ can be linked to the isotropic covariance $C_0$ of $Z$ through the relation
\begin{equation}
\gamma_0(r)= C_0(0)-C_0(r), \quad r\in\R_+ \peq
\end{equation}
For any $r>0$, an estimator of the value of $\gamma_0(r)$ from the observation of a realization $z$ of $Z$ at locations $x_1, \dots, x_{n}\in\mathcal{D}$ is given by the experimental semi-variogram $\widehat{\gamma}_0(r)$, which is defined by:
\begin{equation}
\widehat{\gamma}_0(r)=\frac{1}{2N_{r,\epsilon}}\sum_{\substack{ (i,j)\in\lbrace 1,\dots,n\rbrace^2 \\ r-\epsilon \le\Vert x_i-x_j\Vert\le r +\epsilon }}\left( z(x_i)-z(x_j)\right)^2 \veq
\end{equation}
where $\epsilon>0$ and $N_{r,\epsilon}=\textrm{Card}\left\lbrace (i,j)\in\lbrace 1,\dots,n\rbrace^2 : r-\epsilon \le\Vert x_i-x_j\Vert\le r +\epsilon \right\rbrace$. Here $\epsilon$ is a threshold chosen small in comparison to $r$. Once the experimental semi-variogram has been evaluated for some values of $r>0$,  curve fitting is used in order to choose a function from the catalog of possible isotropic semi-variogram (or equivalently covariance) functions \citep[see e.g.][]{desassis2013automatic}.


\subsection{Geostatistical filtering}

\subsubsection{Presentation of the approach}

Within a geostatistical framework, a noisy observed variable $z$ can be seen as the realization of a Gaussian random field $Z$ defined over a spatial domain $\mathcal{D}$ and that can be decomposed as a sum of $p+1$ independent (Gaussian) random fields $S, N^{(1)}, \dots, N^{(p)}$ ($p\ge 1$), each one characterized by its own covariance model:
\begin{equation}
Z=S+N^{(1)}+\dots+N^{(p)} \peq
\end{equation}
Note in particular that following the independence assumption, the covariance function of $Z$ is simply the sum of the covariance functions of each component forming $Z$.

Filtering then consists in estimating one of these components, here denoted by $Z_s$, from the observation of a realization $z$ of $Z$ at locations $x_1, \dots, x_n\in\mathcal{D}$. In the remaining of this text, we will call true signal the component $S$ that we wish to extract, and noise fields the discarded components $N^{(1)}, \dots, N^{(p)}$.

The factorial kriging method, proposed by \citet{matheron1982pour}, solves this problem by estimating the value of the true signal using the observed noisy signal in a kriging approach \citep{wackernagel2013multivariate}. Formally, the value of the true signal $S$ at some location $x\in\mathcal{D}$ is estimated using a linear combination of the values taken by the noisy signal $Z$ at the observation locations $x_1,\dots,x_n$, which are weighted so that the estimator is both unbiased and minimizes the variance of the resulting error.

 Assuming that all the fields $S,N^{(1)},\dots, N^{(p)}$ are zero-mean,  we get the following estimator $S^*(x)$ for the value of $S$ at $x$:
\begin{equation}
S^*(x)=\left(
\left(\bm\Sigma_S + \bm\Sigma_N^{(1)}+\dots+\bm\Sigma_N^{(p)}\right)^{-1}
\begin{pmatrix}
C_S(x,x_1) \\
\vdots \\
C_S(x,x_1)
\end{pmatrix}
\right)^T
\begin{pmatrix}
Z(x_1) \\
\vdots \\
Z(x_N)
\end{pmatrix} \veq
\label{eq:fkr_def}
\end{equation}
where $C_S$ is the covariance function of $S$, $\bm\Sigma_S$ is the $n\times n$ matrix whose entries are:
\begin{equation}
\left[\bm\Sigma_S\right]_{ij}=C_S(x_i,x_j), \quad 1\le i,j\le n \veq
\end{equation}
and similarly $\bm\Sigma_N^{(1)}, \dots, \bm\Sigma_N^{(p)}$ denote the covariance matrices of respectively $N^{(1)}, \dots, N^{(p)}$ at the same locations. In particular, the estimation of the true signal at the observation locations using now the actual data $z(x_1),\dots,z(x_n)$ can be obtained solving a linear system obtained from the vectorization of \eqref{eq:fkr_def}. Namely, these estimates $s^*(x_1),\dots,s^*(x_n)$ are given by:
\begin{equation}
\begin{pmatrix}
s^*(x_1) \\
\vdots \\
s^*(x_n)
\end{pmatrix}
=
\bm\Sigma_S\left( 
\bm\Sigma_S + \bm\Sigma_N^{(1)}+\dots+\bm\Sigma_N^{(p)}\right)^{-1}
\begin{pmatrix}
z(x_1) \\
\vdots \\
z(x_n)
\end{pmatrix} \peq
\label{eq:syst_filt}
\end{equation}

If we assume that the true signal and all the noise components are second-order stationary and isotropic, the covariance matrices in \eqref{eq:syst_filt} are obtained by applying the corresponding isotropic covariance function to the entries of the (Euclidean) distance matrix of the observation locations $x_1,\dots,x_n\in\mathcal{D}$. These isotropic covariance functions are obtained from the data using the variogram modeling approach described in \Cref{sec:geos_model}. Indeed, the experimental semi-variogram of $Z$ is fitted by a sum of isotropic covariance functions chosen from the catalog. Each one of these functions corresponds in our model to the covariance function of one of the components composing $Z$. It then falls on the user to choose which one of them has to be considered as the model of the true signal, in order to form the covariance matrices and then solve the system \eqref{eq:syst_filt}.

\subsubsection{Limits of the approach}

When dealing with real datasets however, and especially with large seismic datasets, the approach outlined up until now meets two main challenges. First, note that the system \eqref{eq:syst_filt} involves $n\times n$ matrices where $n$ is the number of observation points. Hence a naive approach to solve  \eqref{eq:syst_filt} would require enough storage space to keep in memory the $\mathcal{O}(n^2)$ values of each matrix and about $\mathcal{O}(n^3)$ operations to compute the solution. Clearly, a problem arises when $n$ becomes large, as it is the case for seismic datasets for which $n$ can be equal to several millions of points. 

Nevertheless, under the assumptions of second-order stationarity and isotropy of the random fields, several approaches have been proposed to circumvent this \q{big $n$} problem. We can for instance cite the use of compactly supported \citep{gneiting2002compactly} or tapered  \citep{kaufman2008covariance,furrer2006covariance} covariance functions, which results in sparse\footnote{A sparse matrix a matrix whose most of its entries are zero.} covariance matrices and therefore   a reduction of storage needs and computational costs using algorithms designed for sparse matrices to solve \eqref{eq:syst_filt}. Similarly, imposing that the considered random fields are Markovian ensures that the resulting precision matrices\footnote{A precision matrix is the inverse of a covariance matrix.} are sparse \citep{rue2005gaussian}.  Reformulating \eqref{eq:syst_filt} in terms of precision matrices then allows to once again use sparse matrix algorithms.

Unfortunately, as one may suspect, stationarity and isotropy are strong assumptions that cannot be applied to model any spatial dataset. Dealing with data for which the highly regular spatial structure implied by the isotropic stationary assumption does not apply, requires more work. In the non-stationary case, the covariance function can no longer be expressed as a simple function of the distance between the points, but has an expression that depends on the location and relative position of the considered pair of points. Building and working with the corresponding covariance matrices becomes a complicated task that involves either non-trivial and computationally expensive methods such as Karhunen--Lo\`eve expansions \citep{lindgren2012stationary}, space deformations \citep{sampson1992nonparametric} or convolution models \citep{higdon1999non} (see \citet{fouedjio2017second} for a  complete review); or the use of a restricted class of modeling covariance functions (like for instance the compactly supported non-stationary covariance models proposed by \citet{liang2016class}).

\begin{figure}
\centering
\includegraphics[width=0.5\textwidth]{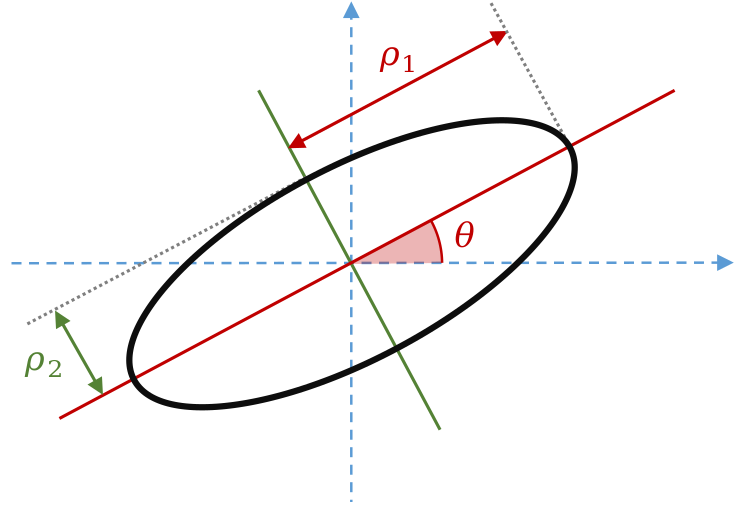}
\caption{Ellipse representation of anisotropy parameters in a two-dimensional domain. The anisotropy angle is $\theta$ and the anisotropy ranges are $\rho_1$ and $\rho_2$.}
\label{fig:ellipse_aniso}
\end{figure}

In the applications considered in this work however, we assume that some prior structural information on the behavior of the random field across the domain is available. Namely, we assume that the random field shows \textit{local anisotropies}: around each point of the domain, there is a preferential direction along which the range of highly correlated values is maximal, whereas it is minimal in the cross-direction(s). In particular, the angles defining the preferential directions are called anisotropy angles and the size of the ranges are called anisotropy ranges. These anisotropy parameters are graphically represented by an ellipse whose axes length and direction are respectively given by the anisotropy ranges and angles (see \Cref{fig:ellipse_aniso} for an example). The goal is then to come up with covariance models/matrices that honor this prior information while keeping the big $n$ problem at bay.

A commonly used solution consists in performing the estimation of the filtered output at a given point considering only a small local neighborhood of observation points around it, and building the resulting (smaller) kriging system while considering the anisotropy parameters constant within the neighborhood \citep{magneron2009m}. However, when applied to large seismic volumes, this method tends to be time consuming and sometimes introduces artifacts that are due to the fact that a collection of small local models are now considered instead of a big global one. This work
aims at presenting an approach to geostatistical filtering that builds on the \q{Stochastic Partial Differential Equations (SPDE) approach} introduced by \citet{lindgren2011explicit}, and that allows filtering to be performed on a global setting while still addressing the limitations of the big $n$ problem.


\section{A matrix-free approach to geostatistical filtering}
\label{sec:mf}


\subsection{Starting point: the SPDE approach}

The SPDE approach, introduced by \citet{lindgren2011explicit} consists in viewing a (Gaussian) random field as
a solution of a class of SPDEs rather than adopting the classical covariance-based approach. It is based on a result from \citet{rozanov1977} that states that any (second-order stationary) isotropic Gaussian Markov random field $Z$ defined over $\R^d$ ($d \ge 1$) is equivalently characterized by:
\begin{itemize}
\item having a spectral density $f_0$ which is the inverse of a polynomial $P_0$ which takes strictly positive values over $\R_+$:
\begin{equation}
\forall \xi\in\R^d, \quad f_0(\Vert\xi\Vert)=\frac{1}{P_0(\Vert\xi\Vert)} \; ;
\end{equation}
\item being a stationary solution of the following SPDE:
\begin{equation}
P_0(-\Delta)^{1/2}Z=W \quad \text{over } \R^d\veq
\label{eq:spde_markov}
\end{equation}
where $P_0(-\Delta)^{1/2}$ is the pseudo-differential operator defined by 
\begin{equation}
P_0(-\Delta)^{1/2}[\cdot]=\mathcal{F}^{-1}\left[\xi \mapsto \sqrt{P_0(\Vert\xi\Vert^2)}\mathcal{F}[\cdot](\xi) \right] \veq
\end{equation}
and $W$ denotes a Gaussian white noise defined over $\R^d$.
\end{itemize}
Example of such fields include fields with a Mat\'ern covariance function\footnote{The Mat\'ern covariance function is a covariance function widely used in geostatistical applications due to its large flexibility. It is defined through three parameters: a range parameter $a>0$ which acts like a scaling parameter, a sill parameter $\sigma^2>0$ which corresponds to the marginal variance of the resulting field, and a shape parameter $\nu>0$ which corresponds to the smoothness of the field. Its expression is: $$C(r)=\frac{\sigma^2}{2^{\nu-1}\Gamma(\nu)}\left(\frac{h}{a}\right)^{\nu}K_{\nu}\left(\frac{h}{a}\right), \quad r\ge 0 \veq$$ where $\Gamma$ is the Gamma function and $K_{\nu}$ is the modified Bessel function of the second kind of order $\nu$.}  with shape parameter $\nu$ that satisfies $\nu+d/2\in\mathbb{N}$ (where $d$ is the dimension of the space on which the field is defined).

The SPDE approach of \citet{lindgren2011explicit} relies on this last characterization. Indeed, they propose to formulate a solution for this SPDE using the finite element method. First, the domain $\mathcal{D}$ is triangulated. The solution of the SPDE  is then approximated by a linear combination of interpolation functions defined over $\mathcal{D}$, and each associated to one of the triangulation nodes. Hence, the solution is written as
\begin{equation}
Z(x)=\sum_{k=1}^{n} z_k \psi_k(x), \quad x\in\mathcal{D} \veq
\label{eq:discr_sol_fem}
\end{equation}
where $n$ denotes the number of triangulation nodes,  $\psi_k$ ($1\le k\le n$) is the interpolating function associated with the $k$-th node, and $z_k$ ($1\le k\le n$) is a Gaussian weight associated with $\psi_k$. A classical choice for the interpolation functions is piecewise linear functions, that are $1$ at a given node and $0$ at any other node. With this choice, the weight $z_k$ actually corresponds to the value of $Z$ at the $k$-th node.

The formulation \eqref{eq:discr_sol_fem} allows the conversion of the SPDE into a linear system involving the random weights $\left\lbrace z_k\right\rbrace_{1\le k\le n}$ , which in turn provides an expression for the covariance (or precision) matrix of these weights. \citet{lindgren2011explicit} actually provide the expression of the precision matrix $\bm Q_Z$ of the random weights $\left(z_1,\dots,z_n\right)$ of the finite element approximation of the solution of \eqref{eq:spde_markov}:
\begin{equation}
\bm Q_Z = \bm C^{1/2}P_0(\bm S)\bm C^{1/2} \veq
\label{eq:prec_mat_markov}
\end{equation}
where $\bm C^{1/2}$ is the diagonal matrix whose entries are
\begin{equation}
[\bm C^{1/2}]_{ii}=\sqrt{\langle \psi_i, 1\rangle}, \quad 1\le i\le n \veq
\label{eq:mat_cdem}
\end{equation}
$\bm S$ is a symmetric positive semi-definite matrix with entries
\begin{equation}
[\bm S]_{ij}=\frac{\langle \nabla\psi_i, \nabla\psi_j\rangle}{\sqrt{\langle \psi_i, 1\rangle \langle \psi_j, 1\rangle}} \quad 1\le i,j\le n \veq
\label{eq:mat_s}
\end{equation}
and $\langle\cdot,\cdot\rangle$ denotes the usual inner product associated with square-integrable functions of $\mathcal{D}$: $\langle \varphi_1,\varphi_2\rangle=\int_{\mathcal{D}}\varphi_1(x)\varphi_2(x)dx$. 

Note that working with the piecewise linear functions described above yields that the matrix $\bm S$ is actually sparse: indeed, $[\bm S]_{ij}$ is zero whenever the supports of the functions $\psi_i$ and $\psi_j$ are disjoint, which is the case whenever the nodes $i$ and $j$ do not belong to a common triangle. Then, the precision matrix in \eqref{eq:prec_mat_markov}, which corresponds to the precision matrix of $Z$ at the triangulation nodes, is a matrix polynomial of a sparse matrix. This means in particular that solving the SPDE using this method actually yields Markovian solutions.

This approach sparked a lot of interest for several reasons. On one hand, the precision matrix of the weights obtained by the SPDE approach being sparse, it provides a practical solution to the big $n$ problem. 

\begin{figure}
\centering
\includegraphics[width=0.7\textwidth]{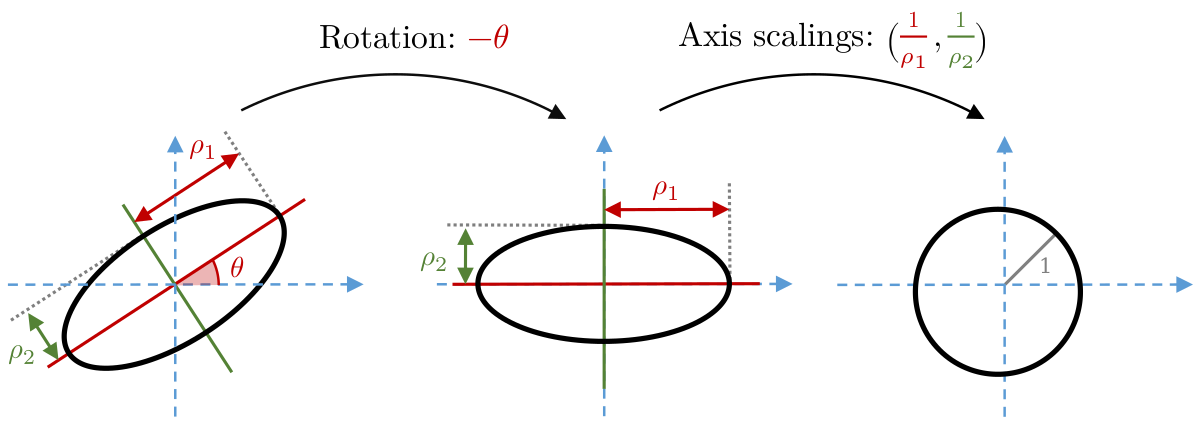}
\caption{Elementary deformations turning anisotropy ellipses into \q{isotropy} circles.}
\label{fig:def_ell}
\end{figure}

On the other hand, adjustments can be made to the model to produce random fields that are both Markovian and non-stationary with locally varying anisotropies. \citet{lindgren2011explicit} and then \citet{fuglstad2015exploring} worked in the particular case where $P_0$ is the polynomial given by $P_0(\lambda)=\frac{1}{\tau}(\kappa^2+x)^2$ for some $\kappa>0$ and $\tau>0$. In this case, the pseudo-differential operator $P_0(-\Delta)^{1/2}$ reduces to a simple differential operator and the the SPDE \eqref{eq:spde_markov} becomes:
\begin{equation}
\kappa^2 Z - \Delta Z = \tau W \quad \text{over } \R^d\peq
\label{eq:spde_markov_alpha2}
\end{equation}
They first propose to work with spatially varying parameters $\kappa$ and $\tau$ in SPDE \eqref{eq:spde_markov_alpha2}, which then creates globally non-stationary fields with a locally isotropic covariance. 
A second approach they suggest consists in defining SPDE \eqref{eq:spde_markov_alpha2} in a deformed space. Rewriting then the SPDE in the original domain using a change of variable yields an expression of the SPDE that is locally parametrized by the Jacobian of the deformation process. This deformation is chosen so that the non-stationary field with given local anisotropies in the original domain becomes a stationary and isotropic field in the deformed domain. In a sense, the ellipses of anisotropy defined on the original domain should become, after this deformation, unit circles (cf. \Cref{fig:def_ell}): hence the deformation acts locally as the composition of a rotation (with angle minus the anisotropy angle) and directional scalings (with factors $1$ over each anisotropy range).

\subsection{An extension of the SPDE approach}

\subsubsection{Laplacian-based random fields}

Several limits to the SPDE approach can be identified. First, it only deals with Markovian random fields, meaning that we are restricted to use only fields whose spectral density can be expressed as the inverse of a polynomial. Besides, it relies on the fact that, given the Markov property, the precision matrix \eqref{eq:prec_mat_markov} should be sparse and therefore can be built and stored. However, the sparsity of the matrix \eqref{eq:prec_mat_markov} is directly linked to the degree of the polynomial $P_0$ defining the spectral density (and the precision matrix itself as it turns out): the larger the degree, the fuller the matrix. Hence, the big $n$ problem might resurface if the degree of the polynomial is not taken small enough. Finally, we note that the  non-stationary setting was outlined only in a particular case of Markovian field, which proves to be an additional limitation for our modeling purposes.

In order to work with non-Markovian spectral densities, \citep{pereiraPhd2019} propose to work with random fields defined using the spectral properties of the Laplace operator. Under the assumption that the domain $\mathcal{D}\subset \R^d$ on which the random fields are defined is bounded (with piecewise smooth boundary), it can be shown that the set $L^2(\mathcal{D})$ of square-integrable functions of $\mathcal{D}$ admits a countable orthonormal basis of eigenfunctions of the Laplacian, i.e functions $\left\lbrace e_k \right\rbrace_{k\in\mathbb{N}}$ satisfying
\begin{equation}
\forall k\in\mathbb{N}, \quad -\Delta e_k = \lambda_k e_k \veq
\label{eq:eig_lap}
\end{equation}
for some $\left\lbrace \lambda_k \right\rbrace_{k\in\mathbb{N}} \subset \R_+$ (called eigenvalues) such that $\lambda_0\le \dots \le \lambda_k\le \dots$, and satisfying either Dirichlet or Neumann boundary conditions (or a mix of them) \citep{gilbarg2015elliptic,laugesen2018spectral}. Hence, any $\varphi\in L^2(\mathcal{D})$ satisfies the following $L^2$-equality:
\begin{equation}
\varphi=\sum_{k\in\mathbb{N}} \langle e_k, \varphi\rangle e_k, \quad \varphi\in L^2(\mathcal{D}) \peq
\label{eq:decomp_l2}
\end{equation}

Now let $f_0:\R_+\rightarrow\R$ be an integrable function, and for instance the spectral density of some isotropic covariance function $C_0$. Following the representation \eqref{eq:decomp_l2} of elements of $L^2(\mathcal{D})$, consider the $L^2(\mathcal{D})$-valued random variable defined by:
\begin{equation}
Z=\sum_{k\in\mathbb{N}} f_0^{1/2}(\lambda_k)w_k e_k \veq
\label{eq:lap_field}
\end{equation}
where $f_0^{1/2}:\R_+\rightarrow\R$ is a function such that $\left(f_0^{1/2}\right)^2=f_0$ on $\R_+$, $\lbrace\lambda_k\rbrace_{k\in\mathbb{N}}$ is defined in \eqref{eq:eig_lap}, and $\lbrace w_k\rbrace_{k\in\mathbb{N}}$ is a sequence of independent standard Gaussian variables. In particular, it is straightforward to show that then, $Z$ defines a Gaussian random field over $\mathcal{D}$ with covariance function $C_Z$ given by:
\begin{equation}
C_Z(x_1,x_2)=\sum_{k\in\mathbb{N}}f_0(\lambda_k)e_k(x_1)e_k(x_2), \quad x_1,x_2\in\mathcal{D} \peq
\label{eq:cov_lap_field}
\end{equation}
\citet{solin2019hilbert} show that, as defined by \eqref{eq:lap_field}, $Z$ actually approximates a Gaussian random field on $\mathcal{D}$ with (radial) spectral density $f_0$. They even provide a uniform bound on the error between the actual covariance function of $Z$ and the covariance function associated with $f_0$ which proved that the approximation improves as we move further away from the boundary.

\subsubsection{Numerical approximation of the resulting fields}

Building on the SPDE approach presented earlier, \citet{pereiraPhd2019} then proposes to build a finite element approximation\footnote{Or rather a Ritz--Galerkin approximation \citep{strang1973analysis,brenner2007mathematical}.} of the field $Z$ defined by \eqref{eq:lap_field}. Basically, $Z$ is once again replaced by a linear combination of interpolation functions (related to a triangulation of $\mathcal{D}$) where the weights $(z_1, \dots, z_n)$ are chosen so that the approximation coincides with a definition of $Z$ through \eqref{eq:lap_field}, but replacing now the eigenfunctions $\lbrace e_k\rbrace_{k\in\mathbb{N}}$ and eigenvalues $\lbrace \lambda_k\rbrace_{k\in\mathbb{N}}$ of the Laplacian by those of a discretized version of the Laplacian\footnote{In particular, this discretized Laplacian $-\Delta_n$ is defined for any $\varphi\in V_n=\spn{\psi_1,\dots\psi_n}$ by: $$\forall u\in V_n, \quad\langle -\Delta_n f, u\rangle=\langle -\Delta f, u\rangle \peq$$}. \citet[Theorem 7.3.2]{pereiraPhd2019} actually proves that then, these weights are Gaussian, zero-mean and have covariance matrix $\bm\Sigma_Z$ given by:
\begin{equation}
\bm\Sigma_Z=\bm C^{-1/2}f_0(\bm S)\bm C^{-1/2} \veq
\label{eq:cov_mat_weights}
\end{equation}
where $\bm C^{-1/2}$ is the inverse of the matrix $\bm C^{1/2}$ introduced in \eqref{eq:mat_cdem}, $\bm S$ is once gain defined by \eqref{eq:mat_s} and $f_0(\bm S)$ is a matrix function defined through an eigendecomposition of $\bm S$ as
\begin{equation}
\bm S=\bm V
\begin{pmatrix}
\mu_1 & &\\
 & \ddots & \\
 & & \mu_n
\end{pmatrix}
\bm V^T
\Longrightarrow f_0(\bm S)=\bm V
\begin{pmatrix}
f_0(\mu_1) & &\\
 & \ddots & \\
 & & f_0(\mu_n)
\end{pmatrix}
\bm V^T \veq
\label{eq:def_mat_func}
\end{equation}
where $\bm V$ is a matrix whose column are an orthonormal basis of $\R^n$, i.e. $\bm V^{-1}=\bm V^T$, and $\mu_1,\dots,\mu_n\in\R_+$ are eigenvalues of $\bm S$. A convergence result of this approximation towards \eqref{eq:lap_field} as the mesh size of the triangulation decreases is actually provided by \citet[Theorem 8.2.1]{pereiraPhd2019}, thus justifying this approximation approach.

Note that in the case where $f_0$ is a Markovian spectral density, we retrieve exactly the expression \eqref{eq:prec_mat_markov} obtained from the SPDE approach by inverting \eqref{eq:cov_mat_weights}, which bridges the gap between both approaches. The approach proposed by \citet{pereiraPhd2019} can therefore be seen as a generalization of the SPDE approach to any (radial) spectral density. In what follows we show how this new approach can easily be extended to the non-stationary case (without a restriction on the possible models) and how it paves the way for a \q{matrix-free} approach of Geostatistics that pushes the limits of the big $n$ problem.


\subsection{Working with non-stationary models}

The extension of this Laplacian-based model for stationary (isotropic) Gaussian random fields to non-stationary ones relies on the notion of Riemannian manifold \citep{jost2008riemannian}. A Riemannian manifold is the association of a manifold  with a Riemannian metric. On one hand, the manifold is a set that can locally be considered as Euclidean. On the other
hand, the Riemannian metric is an application that smoothly associates to each point $x$ of the manifold an inner product that redefines the notions of length and of angles for infinitely small vectors that would be attached to $x$.

An equivalent of the usual Laplacian that takes into account both the structure of the manifold and the local redefinition of its geometry due to the metric can be defined, and is called Laplace--Beltrami operator \citep{lablee2015spectral}. When considering compact Riemannian manifolds, a spectral theorem similar to the one presented in the previous section can be stated, then yielding a representation of square-integrable\footnote{It is important to note that now, integrals also account for the metric through an infinitesimal integration volume computed in the metric-defined geometry (which varies across the manifold).} functions defined over the manifold in a basis of eigenfunctions of the Laplace--Beltrami operator \citep{lablee2015spectral,jost2008riemannian}. Hence, Gaussian fields can once again be defined by \eqref{eq:lap_field} by using this time the eigenfunctions and eigenvalues of the Laplace--Beltrami operator.

Going back to our initial problem of defining Gaussian fields with local anisotropies on a bounded domain $\mathcal{D}$, \citet{pereiraPhd2019} proposes to work with a Gaussian field defined on a problem-specific Riemannian manifold. Namely, $\mathcal{D}$ is the manifold and a metric is defined so that at each point $x\in\mathcal{D}$, the resulting inner product is the Euclidean inner product between vectors deformed by the composition of rotation and axis scaling introduced in \Cref{fig:def_ell}, and defined for the anisotropy parameters at $x$. In other words, the metric is chosen so that it locally deforms $\mathcal{D}$ into a domain where the anisotropy becomes isotropy. 

Defining Gaussian fields through \eqref{eq:lap_field} using the Laplace--Beltrami operator associated with this Riemannian manifold then yields (away from boundary) Gaussian fields on $\mathcal{D}$ whose covariance function is isotropic after a change of variables from $\mathcal{D}$ to the locally deformed domain described above. Here, once again, the isotropic covariance in question is the covariance function associated with the spectral density $f_0$ used to scale the Gaussian weights in \eqref{eq:lap_field}. Hence we have,
\begin{equation}
C_Z(x, x+h)\approx C_0\left(\Vert \bm D(x)\bm R(x)^T h\Vert\right), \quad 
\begin{cases}
x,x+h\in\mathcal{D} \\ 
h\rightarrow 0 
\end{cases}\veq
\end{equation}
where $\bm D(x)\in\R^{d\times d}$ is the diagonal matrix whose entries are the inverse of the anisotropy ranges at $x$, $\bm R(x)$ is the rotation matrix whose angles are the anisotropy angles at $x$, and $C_0$ is linked to $f_0$ through \eqref{eq:cov_radial}. It is then straightforward to check that such a structure of covariance locally reproduces the anisotropic structure we were aiming at. 

\citet{pereiraPhd2019} carries out the same  finite element approximation scheme  as the one described in the stationary case and end up once again with an explicit characterization of the weights $(z_1,\dots,z_n)$ in \eqref{eq:discr_sol_fem}. For any $x\in\mathcal{D}$, denote by $\rho_1(x),\dots,\rho_d(x) >0$ the anisotropy ranges at $x$ and by $\bm R(x)\in\R^{d\times d}$ the rotation matrix associated with the anisotropy angles at $x$. Introduce then $\bm H : \mathcal{D} \rightarrow \R^{d\times d}$ and $h: \mathcal{D}\rightarrow \R_+$ defined as:
\begin{equation}
\bm H(x)=\bm R(x) \begin{pmatrix}
\rho_1(x)^2 & &\\
& \ddots &\\
& & \rho_d(x)^2
\end{pmatrix}\bm R(x)^T, \quad x\in\mathcal{D} \veq
\end{equation}
and 
\begin{equation}
h(x)=\sqrt{\det \bm H^{-1}(x)}=\prod_{i=1}^d\frac{1}{\rho_i(x)}, \quad x\in\mathcal{D} \peq
\end{equation}
Then \citet[Theorem 7.3.2]{pereiraPhd2019} shows that the approximation weights are once again Gaussian, zero-mean and have covariance matrix $\bm\Sigma_Z$ given by:
\begin{equation}
\bm\Sigma_Z=\tilde{\bm C}^{-1/2}f_0(\tilde{\bm S})\tilde{\bm C}^{-1/2} \veq
\label{eq:cov_mat_weights_aniso}
\end{equation}
where now, $\tilde{\bm C}^{-1/2}$ is the diagonal matrix whose entries are
\begin{equation}
[\tilde{\bm C}^{-1/2}]_{ii}=\frac{1}{\sqrt{\langle \psi_i, h\rangle}}, \quad 1\le i\le n \veq
\label{eq:mat_cdem_a}
\end{equation}
$\tilde{\bm S}$ is a symmetric positive semi-definite matrix with entries
\begin{equation}
[\tilde{\bm S}]_{ij}=\frac{\langle \nabla\psi_i, h\cdot\bm H\nabla\psi_j\rangle}{\sqrt{\langle \psi_i, h\rangle \langle \psi_j, h\rangle}} \quad 1\le i,j\le n \veq
\label{eq:mat_s_a}
\end{equation}
and $\langle\cdot,\cdot\rangle$ denotes once again the usual inner product associated with square-integrable functions of $\mathcal{D}$. Note in particular that \eqref{eq:cov_mat_weights} and \eqref{eq:cov_mat_weights_aniso} actually coincide whenever $\forall x\in\mathcal{D}$, $\rho_1(x)=\dots=\rho_d(x)=1$, which corresponds to isotropic case (all the anisotropy ellipses are unit circles with same radius).


\subsection{A matrix-free implementation}

\subsubsection{The matrix-free approach}

Considering \eqref{eq:cov_mat_weights} and \eqref{eq:cov_mat_weights_aniso}, building the covariance matrices involved for instance to solve the filtering system \eqref{eq:syst_filt} seems like a task particularly impacted by the big $n$ problem. Indeed, their constructions seem to rely on the full eigen decomposition of a (sparse) matrices of size $n$, and results in generally full $n\times n$ matrices. However, the fact that both \eqref{eq:cov_mat_weights} and \eqref{eq:cov_mat_weights_aniso} are expressed as a product of diagonal matrices with a matrix function points towards adopting a so-called matrix-free strategy aiming at working with the covariance matrices without actually building and storing them.

Indeed, getting back to \eqref{eq:syst_filt}, note that the filtered solution can be obtained in two steps:
\begin{itemize}
\item First, solve for $\bm y\in\R^n$ the linear system
\begin{equation}
\left( 
\bm\Sigma_S + \bm\Sigma_N^{(1)}+\dots+\bm\Sigma_N^{(p)}\right)
\bm y
=
\begin{pmatrix}
z(x_1) \\
\vdots \\
z(x_n)
\end{pmatrix} \peq
\label{eq:syst_lin_filt}
\end{equation}
\item Then, return the solution as
\begin{equation}
\begin{pmatrix}
s^*(x_1) \\
\vdots \\
s^*(x_n)
\end{pmatrix}
=
\bm\Sigma_S\bm y \peq
\end{equation}
\end{itemize}

There exists iterative algorithms that allow to solve linear systems while relying only on products between the matrix defining the linear system and vectors \citep{nocedal2006numerical}. In particular, such algorithms can be used in a matrix-free way in the sense that they are able to solve the linear system without needing to explicitly build and store the associated matrix: all that is required is a routine that performs the product between this matrix and an input vector. An example of such an algorithm is the Conjugate Gradient (CG) algorithm \citep{nocedal2006numerical}. Hence, using the CG algorithm to solve \eqref{eq:syst_lin_filt}, we can completely solve the filtering problem \eqref{eq:syst_filt} in a matrix-free way: all that is required is routines computing the product between any one of the matrices $\bm\Sigma_S,\bm\Sigma_N^{(1)},\dots,\bm\Sigma_N^{(p)}$ and an input vector. We present in Algorithm \ref{alg:cg} this approach to geostatistical filtering.

\begin{algorithm}
\begin{minipage}{0.92\textwidth}
\begin{description}
\item[\textbf{Input}] $ $ \\
$\bullet$ A vector of noisy observations $\bm b=(z(x_1),\dots,z(x_n))^T\in \R^n$ of a signal at locations $x_1,\dots,x_n$ of a spatial domain.\\
$\bullet$ A routine $\Pi_{S}(\cdot)$ (resp. $\Pi_{N}^{(1)}(\cdot),\dots,\Pi_{N}^{(p)}(\cdot)$) that returns for any $\bm v\in\R^{n}$ the vector $\bm\Sigma_S\bm v$ (resp. $\bm\Sigma_N^{(1)}\bm v,\dots,\bm\Sigma_N^{(p)}\bm v$) where $\bm\Sigma_S$ (resp. $\bm\Sigma_N^{(1)},\dots,\bm\Sigma_N^{(p)}$) is the covariance matrix of the true signal (resp. of the noise components) at the observation locations. \\
$\bullet$ An initial guess $\bm y^{(0)}\in\R^n$.\\
$\bullet$ A convergence threshold $\tau>0$.
\item[\textbf{Output}] $ $ \\
$\bullet$ The vector of factorial kriging estimates $(s^*(x_1),\dots,s^*(x_n))^T\in\R^n$ of the true signal at the observation locations.\\
\end{description} 
\end{minipage} \\
 \dotfill \\
 	$k=0$ \;
 	$\bm r^{(0)}=\bm b-\left(\Pi_{S}(\bm y^{(0)})+\Pi_{N}^{(1)}(\bm y^{(0)})+\dots+\Pi_{N}^{(p)}(\bm y^{(0)})\right)$ \; 
 	$\bm d^{(0)}=\bm r^{(0)}$ \;
 	$\bm p^{(0)}=\Pi_{S}(\bm d^{(0)})+\Pi_{N}^{(1)}(\bm d^{(0)})+\dots+\Pi_{N}^{(p)}(\bm d^{(0)})$\;
 	\While{$\left(\Vert \bm r^{(k)}\Vert > \tau\right)$}{
 	$\alpha_k=\frac{\left(\bm r^{(k)}\right)^T\bm r^{(k)}}{\left(\bm d^{(k)}\right)^T\bm p^{(k)}}$ \;
 	$\bm y^{(k+1)} =\bm y^{(k)}+\alpha_k \bm d^{(k)}$ \;
 	$\bm r^{(k+1)} =\bm r^{(k)}-\alpha_k\bm p^{(k)}$\;
 	$\beta_k=\frac{\left(\bm r^{(k+1)}\right)^T\bm r^{(k+1)}}{\left(\bm r^{(k)}\right)^T\bm r^{(k)}}$\;
 	$\bm d^{(k+1)}=\bm r^{(k+1)}+\beta_k\bm d^{(k)}$\;
 	$\bm p^{(k+1)}=\Pi_{S}(\bm d^{(k+1)})+\Pi_{N}^{(1)}(\bm d^{(k+1)})+\dots+\Pi_{N}^{(p)}(\bm d^{(k+1)})$\;
 	$k\leftarrow k+1$\;
 	}
 \Return{$\Pi_{S}(\bm y^{(k)})$}. 
 \caption{Matrix-free Conjugate Gradient for geostatistical filtering.}\label{alg:cg}
\end{algorithm}

Apart from the routines, note that each iteration of Algorithm \ref{alg:cg} requires at most $\mathcal{O}(n)$ operations and memory space. Hence the computational and memory efficiency of the algorithm comes down to that of the routines used to compute the products between one of the covariance matrices and vectors. To achieve it, we propose to model the true signal and the noise components using the approach presented in the previous section. Hence, every covariance matrix in \eqref{eq:syst_filt} can be written as \eqref{eq:cov_mat_weights_aniso} with specific matrices $\tilde{\bm C}^{-1/2}$ and $\tilde{\bm S}$ obtained after a triangulation of the spatial domain and taking into account the local anisotropy parameters of the corresponding field. Then we use a polynomial trick to efficiently compute the product between these covariance matrices and vectors. Such an approach has been proposed by  \citet{Dietrich1995} and \citet{pereira2019efficient} to deal with the simulation of Gaussian random fields, and by for instance \citet{higham2008functions} and \citet{hammond2011wavelets} for dealing with applications involving matrix functions.

\subsubsection{Polynomial trick for matrix-vector products}

A product between a covariance matrix of the form \eqref{eq:cov_mat_weights_aniso} and a vector $\bm v\in\R^n$ is obtained in three steps:
\begin{enumerate}
\item Compute the product $\bm u_1=\tilde{\bm C}^{-1/2}\bm v$.
\item Compute the product $\bm u_2=f_0(\tilde{\bm S})\bm u_1$.
\item Return the product $\tilde{\bm C}^{-1/2}\bm u_2$.
\end{enumerate}
On one hand, steps 1 and 3 consists in products between a diagonal matrix and a vector, and hence require $\mathcal{O}(n)$ operations if the diagonal entries of $\tilde{\bm C}^{-1/2}$ are stored. On the other, step 2 would in general require either to have diagonalized the matrix $\bm S$ at some point and to have stores its full set of eigenvectors and eigenvalues.

 However, whenever $f_0$ is a polynomial $f_0(\lambda)=\sum_{k=0}^Ka_k\lambda^k$, it is straightforward to check (via a proof by induction) that the definition of $f_0(\tilde{\bm S})$ through \eqref{eq:def_mat_func} is equivalent to the traditional notion of matrix polynomial, i.e.
\begin{equation}
f_0(\tilde{\bm S})=\sum_{k=0}^K a_k\tilde{\bm S}^k \veq
\end{equation}
where $\bm S^0$ is the identity matrix. The product of step 2 can then be computed without requiring any diagonalization as its given by 
\begin{equation}
\begin{aligned}
\bm u_2&=\sum_{k=0}^K a_k\tilde{\bm S}^k\bm u_1\\
&=a_0\bm u_1+\tilde{\bm S}\cdot\left( a_1\bm u_1 +\tilde{\bm S}\cdot\left( \cdots +\tilde{\bm S}\cdot\left( a_{K-1}\bm u_1 + a_K\tilde{\bm S}\cdot \bm u_1 \right)\dots\right)\right) \veq
\end{aligned}
\end{equation}
and therefore can be computed with an iterative approach that requires at each iteration only a single product between $\bm S$ and a vector. Hence, in the polynomial case, only $\bm S$ needs to be stored which represents a storage need of $\mathcal{O}(m_{\mathtt{nz}}n)$ where $m_{\mathtt{nz}}\ll n$ is the mean number of non-zero entries on a row of $\tilde{\bm S}$, and the computational cost of step 2 now amounts to $\mathcal{O}(Km_{\mathtt{nz}}n)$ operations.

Now, when $f_0$ is not polynomial, we propose to replace it by an appropriate polynomial $P_{f_0}$. In particular, this polynomial is chosen so that it is a good approximation of $f_0$ over a segment $[0,l]$ that contains the $n$ eigenvalues of $\tilde{\bm S}$. This is sufficient to ensure that then $f_0(\tilde{\bm S})\bm u_1\approx P_{f_0}(\tilde{\bm S})\bm u_1$, following the definition \eqref{eq:def_mat_func} of matrix functions and the fact that by definition of $P_{f_0}$, $\forall i\in\lbrace 1,\dots,n\rbrace$, $P_{f_0}(\lambda_i)\approx f_0(\lambda_i)$. As for the segment of approximation, it can be obtained by upper-bounding the largest eigenvalue of $\tilde{\bm S}$, which can be done at a negligible cost (of order $\mathcal{O}(m_{\mathtt{nz}}n)$) using either the Gershgorin circle theorem \citep{gershgorin1931uber} or the fact that
\begin{equation}
\sqrt{\textrm{Trace}\left(\tilde{\bm S}\tilde{\bm S}^T\right)}
=\sqrt{\sum_{i=1}^n\sum_{j=1}^n [\tilde{\bm S}]_{ij}^2}
=\sqrt{\sum_{i=1}^n\lambda_i^2}\ge \max_{1\le i\le n}\lambda_i \peq
\label{eq:max_eig}
\end{equation}

In practice, the polynomial $P_{f_0}$ is determined using Chebyshev polynomial approximation \citep{mason2002chebyshev,press2007numerical}. As outlined by \citet{pereira2019efficient}, this choice guarantees (for $f_0$ Lipschitz-continuous):
\begin{itemize}
\item the uniform convergence of the approximation over the segment as the degree of the chosen polynomial increases,
\item the fact that at any order of approximation the polynomial is close (with respect to the uniform norm) to the best polynomial approximation of same degree,
\item the fact that the coefficients of the polynomial in the Chebyshev basis of polynomials can be computed very efficiently using the Fast Fourier Transform (FFT) algorithm of \citet{cooley1965algorithm}, which has a complexity of $\mathcal{O}(K\log K)$ for the computation of $K$ coefficients.
\end{itemize}
As for the choice of the order of approximation, \citet{pereira2019efficient} provide heuristics based on the theory of statistical tests and on the explicit derivation of numerical errors.

We now have all the ingredients to produce the missing piece of our matrix-free approach to filtering: computationally efficient routines to compute the product between covariance matrices of the form \eqref{eq:cov_mat_weights_aniso} and vectors. These are laid out in Algorithm \ref{alg:prod}. Note in particular that the storage needs of this type of routine are order $\mathcal{O}(m_{\mathtt{nz}} n)$ and its computational cost is of order $\mathcal{O}(Km_{\mathtt{nz}} n)$ (at each iteration, the most costly operation is the product between $\tilde{\bm S}$ and a vector). Hence these routines are highly scalable given that they scale linearly with the size $n$ of the problem.

\begin{algorithm}
\begin{minipage}{0.92\textwidth}
\begin{description}
\item[\textbf{In storage}] $ $ \\
$\bullet$ The matrices $\tilde{\bm C}^{-1/2}$ and $\tilde{\bm S}$.\\
$\bullet$ A segment $[0,l]$ containing all the eigenvalues of $\tilde{\bm S}$.\\
$\bullet$ An approximation order $K\in\mathbb{N}$.\\
$\bullet$ The coefficients $c_0,\dots,c_K\in\R$ of the Chebyshev polynomial approximation of $f_0$ over $[0,l]$ (in the Chebyshev polynomial basis).
\item[\textbf{Input}] $ $ \\
$\bullet$ A vector $\bm v\in\R^n$.
\item[\textbf{Output}] $ $ \\
$\bullet$ (An approximation of) the vector $\tilde{\bm C}^{-1/2}f_0(\tilde{\bm S})\tilde{\bm C}^{-1/2}\bm v$.  \\
\end{description} 
\end{minipage} \\
 \dotfill \\
  $\bm x =\bm \tilde{\bm C}^{-1/2}\bm v$ \;
 $\bm u^{(-2)}= \bm u^{(-1)}= \bm u= \bm y=\bm 0$\;
  \For{$k$ \KwFrom $0$ \KwTo $K$}{
	\uIf{$k=0$}{
    $\bm u \leftarrow \frac{1}{2}\bm x$ \;
  }
  \uElseIf{$k=1$}{
    $\bm u \leftarrow \frac{2}{l}\bm S\cdot \bm x-\bm x$
  }
  \Else{
    $\bm u \leftarrow \frac{4}{l}\bm S\cdot\bm u^{(-1)}-2\bm u^{(-1)}-\bm u^{(-2)}$ \;
  }  
   	$\bm y \leftarrow \bm y + c_k \bm u$ \;
  	$\bm u^{(-2)} \leftarrow \bm u^{(-1)}$ \;
  	$\bm u^{(-1)} \leftarrow \bm u$ \;
 }
 \Return{ $\bm \tilde{\bm C}^{-1/2}\bm y$}.
 \caption{Routine to compute the product between a covariance matrix of the form \eqref{eq:cov_mat_weights_aniso} and an input vector.}\label{alg:prod}
\end{algorithm}

\subsubsection{Summary of the proposed implementation}

\begin{figure}
\centering
\includegraphics[width=\textwidth]{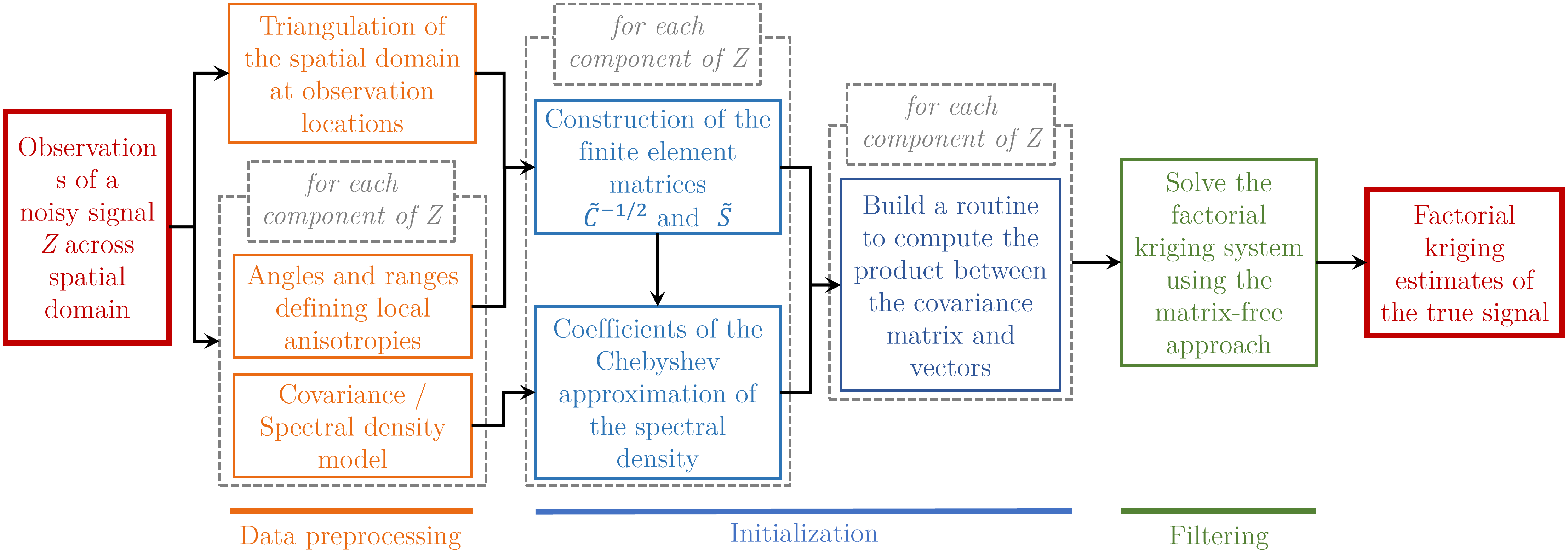}
\caption{Matrix-free approach to geostatistical filtering.}
\label{fig:pres_filt}
\end{figure}

We now conclude the presentation of matrix-free approach to geostatistical filter that we propose, and that we summarized in \Cref{fig:pres_filt}. Starting from a noisy signal observed across a spatial domain, we first determine the anisotropy and covariance parameters characterizing each component of the signal. The anisotropy parameters can be determined using for instance gradient based algorithms on a smoothed out version of the signal, or using auxiliary variables. The covariance parameters can be determined using local variogram modeling. This first step is actually common to any geostationary filtering approach aiming at incorporating local information into the process \citep{magneron2009m,piazza2015m}. Besides, the spatial domain is triangulated at the observed locations. In the particular case of observation on a grid, the triangulation needs no work as we can simply divide the grid cells into triangles or tetrahedrons.

Then, for each component of the signal, the corresponding finite element matrices $\tilde{\bm C}^{-1/2}$ and $\tilde{\bm S}$, which incorporate the local anisotropy information, are built using \eqref{eq:mat_cdem_a} and \eqref{eq:mat_s_a}. An interval $[0,l]$ containing all the eigenvalues of $\tilde{\bm S}$ is deduced (for instance through \eqref{eq:max_eig}) and the coefficients of the Chebyshev polynomial approximation of the spectral spectral density of the component over $[0,l]$ are computed. Then, for each component, the routines of Algorithm \ref{alg:prod} aiming at computing the product between the covariance matrix of the component and vectors are written.  Finally, Algorithm \ref{alg:cg} is used to compute the factorial kriging estimates of the true signal.

\section{Applications}
\label{sec:appli}


\subsection{Synthetic case study}

\begin{figure}[t]
     \centering
    \begin{subfigure}[t]{0.32\textwidth}
    	\centering
        \includegraphics[width=\textwidth]{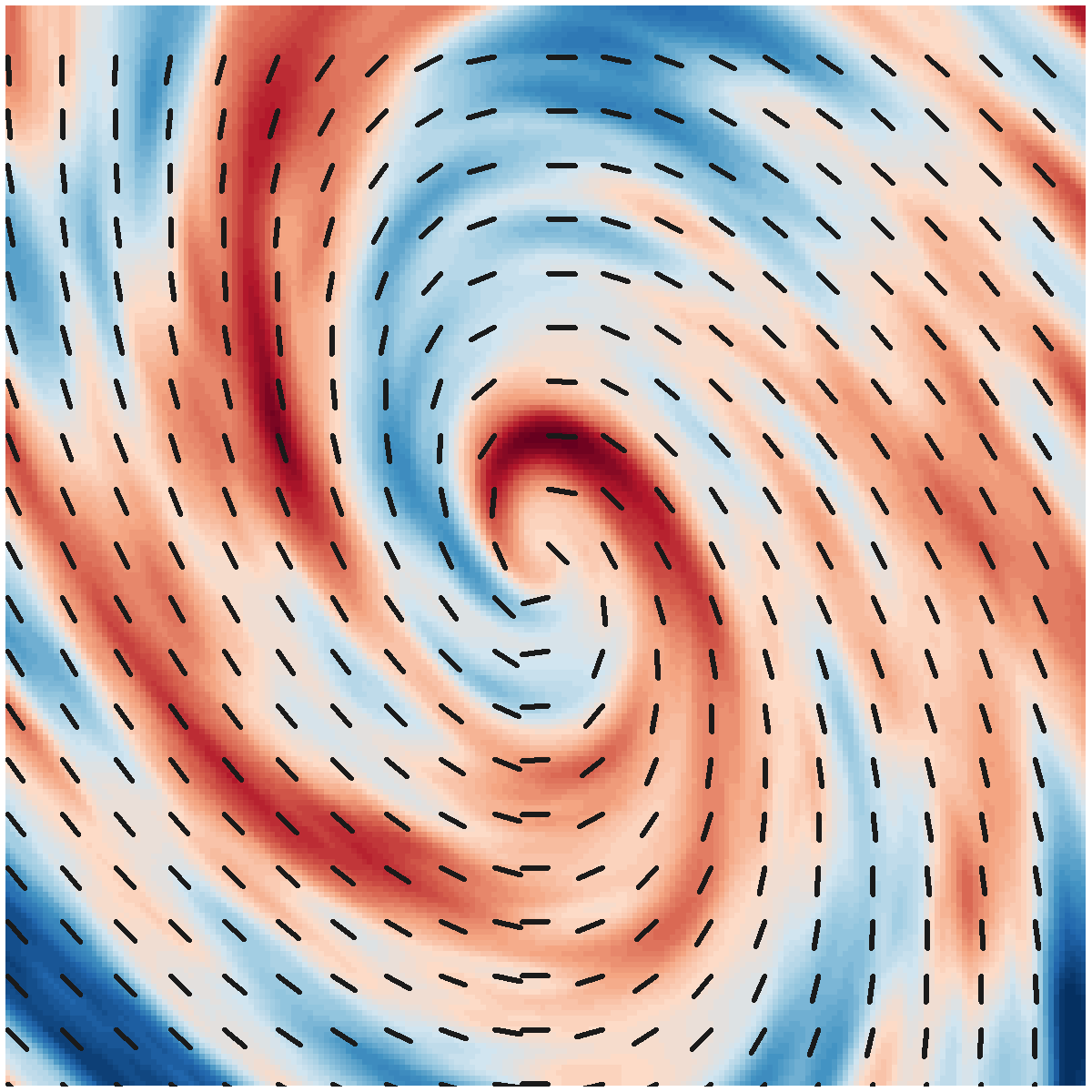}
        \caption{Simulated \q{signal} and associated local anisotropies.}
    \end{subfigure}
    \begin{subfigure}[t]{0.32\textwidth}
    	\centering
        \includegraphics[width=\textwidth]{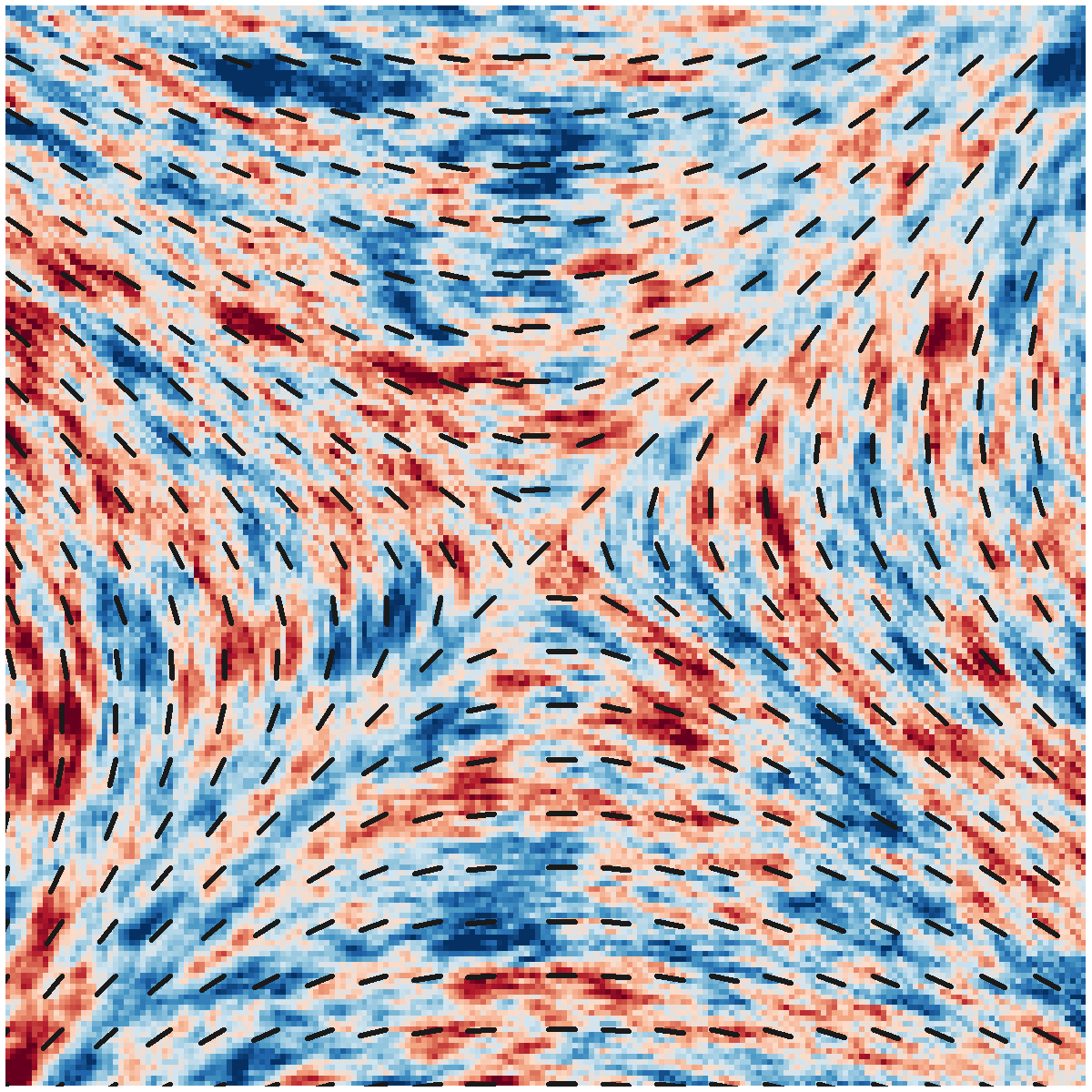}
        \caption{Simulated \q{noise} and associated local anisotropies.}
    \end{subfigure}
        \begin{subfigure}[t]{0.32\textwidth}
    	\centering
        \includegraphics[width=\textwidth]{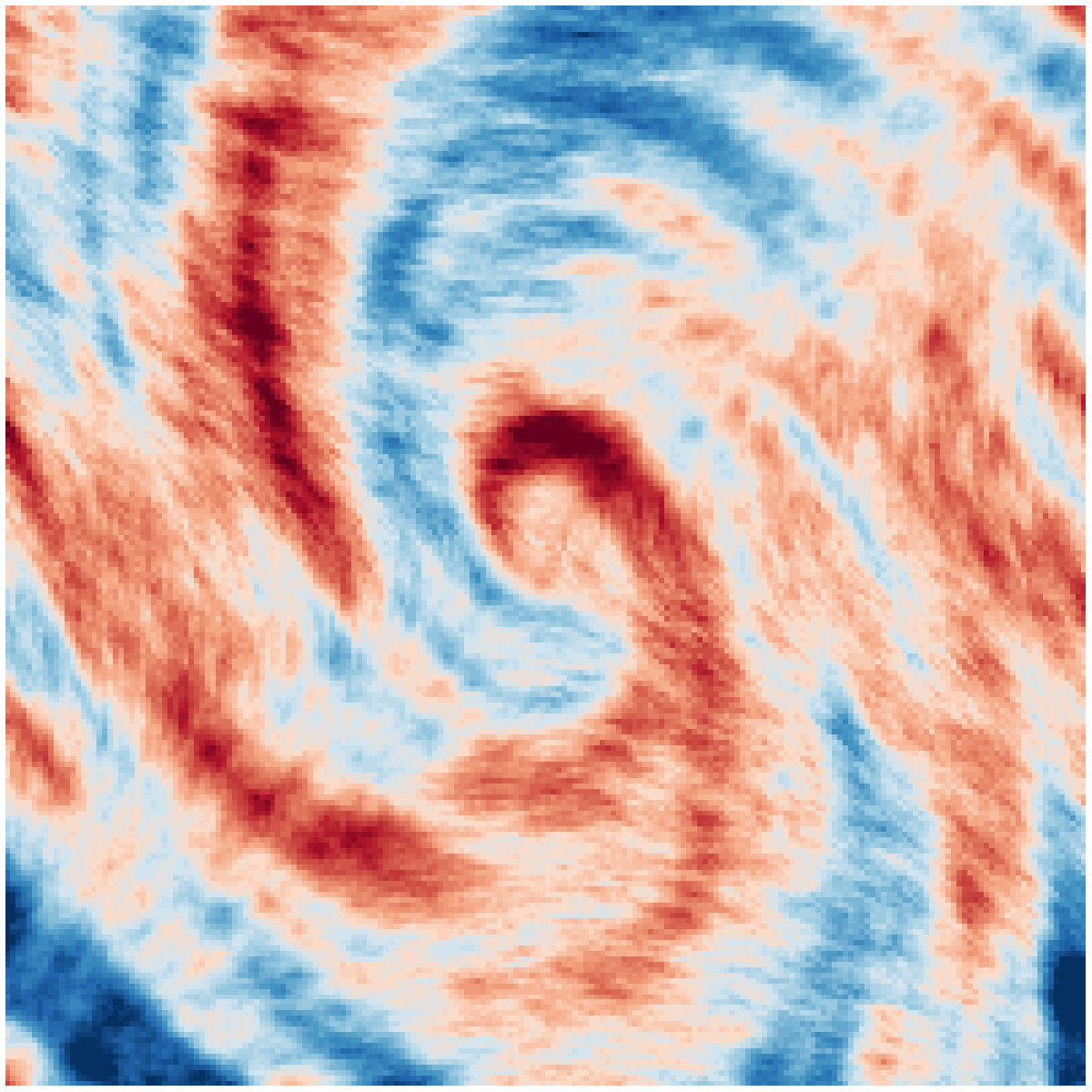}
        \caption{Noisy signal.}
    \end{subfigure}
        \caption{Simulated data for the filtering test. The noisy signal (c) is the sum of the simulated \q{signal} (a) and the simulated \q{noise} (b). }
    \label{fig:filt_sim}
\end{figure}

In this first case study, we simulated two Gaussian random fields on a $400\times 400$ grid, both presenting local anisotropies. We summed them to define our input (noisy) signal (cf. \Cref{fig:filt_sim}). Hence, the noisy signal to be filtered is composed of:
\begin{itemize}
\item A non-stationary field defined by a Mat\'ern covariance function with smoothness parameter $3$, ranges $100$ and $20$ along its principal directions, and sill $1$. It has local anisotropies that describe a vortex-like shape. This field is the true signal we want to extract.
\item A non-stationary field defined by an exponential covariance function with ranges $25$ and $8$ along its principal directions, and sill $0.4$. It has local anisotropies that describe a \q{X} shape.
\end{itemize}

\begin{figure}
     \centering
    \begin{subfigure}[t]{0.32\textwidth}
    	\centering
        \includegraphics[width=\textwidth]{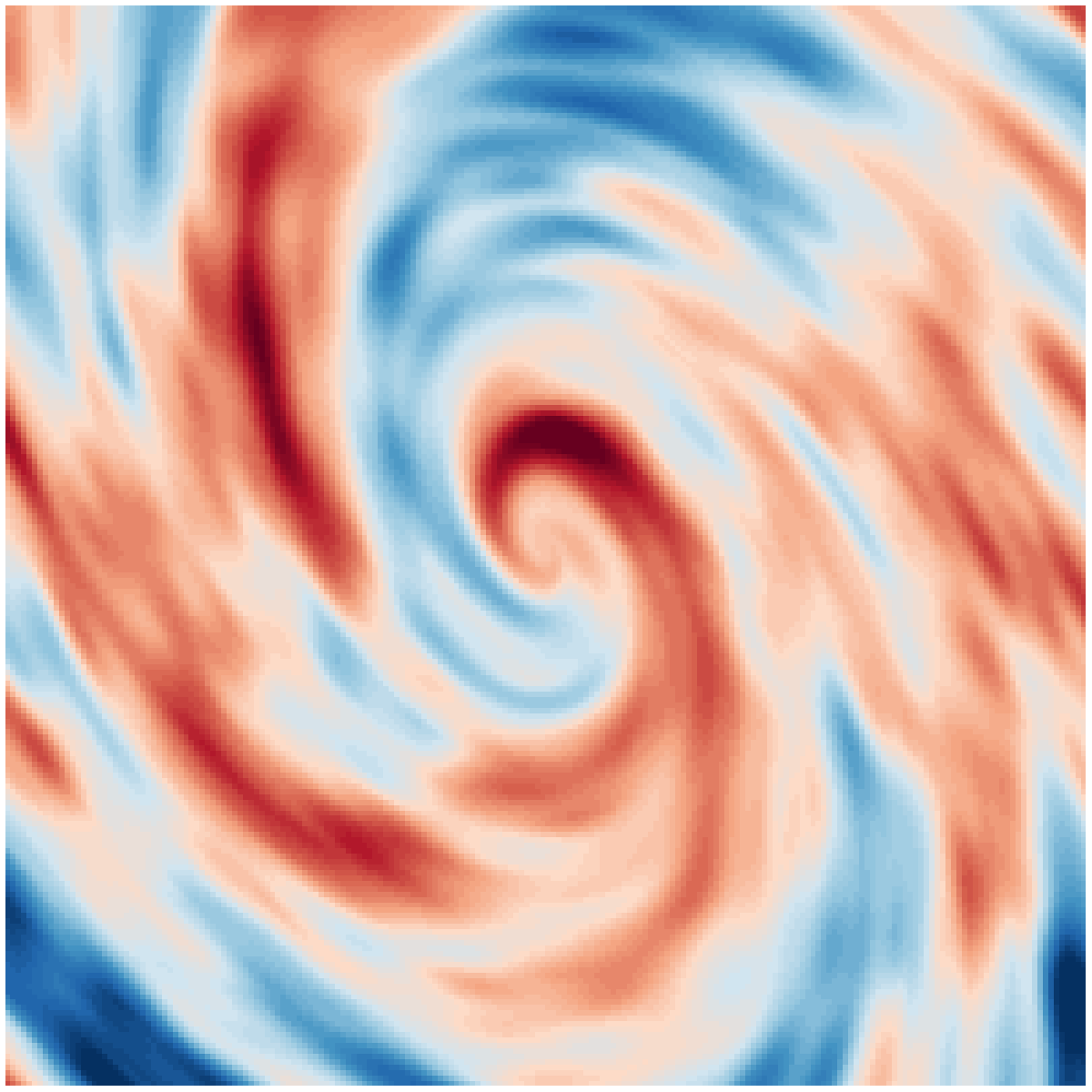}
        \caption{Filtered output.}
    \end{subfigure}
    \begin{subfigure}[t]{0.32\textwidth}
    	\centering
        \includegraphics[width=\textwidth]{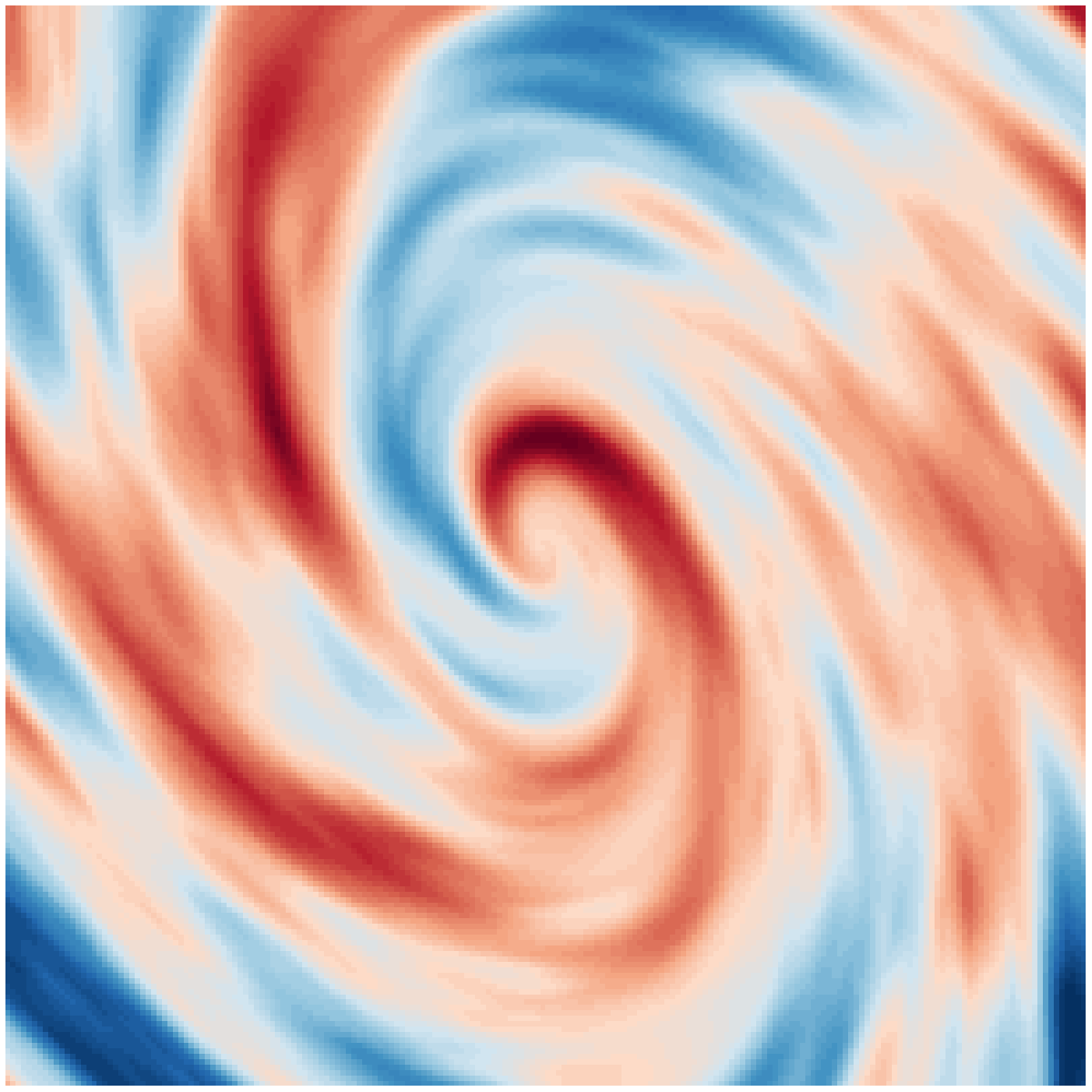}
        \caption{Original signal.}
    \end{subfigure}
        \begin{subfigure}[t]{0.32\textwidth}
    	\centering
        \includegraphics[width=\textwidth]{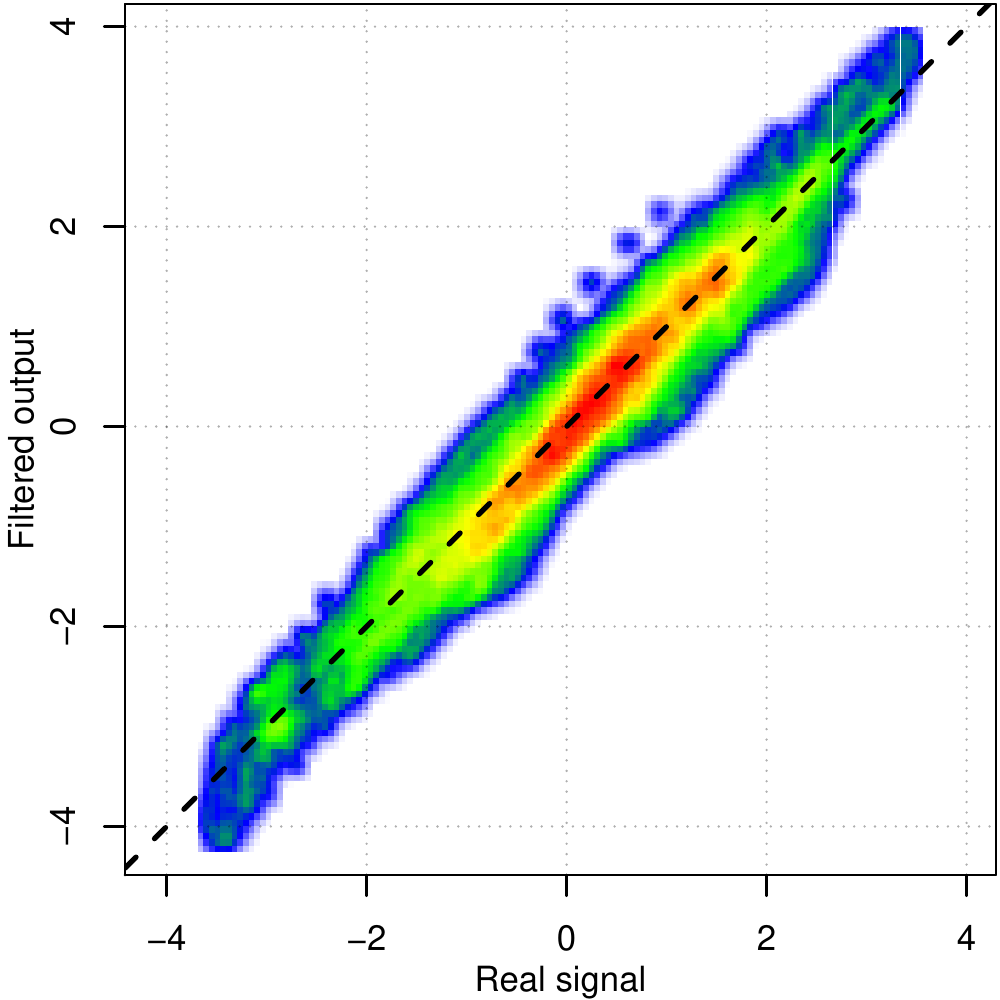}
        \caption{Density scatter plot.}
    \end{subfigure}
        \caption{Results of the filtering algorithm applied to the simulated data.}
    \label{fig:filt_res}
\end{figure}

The filtering process is launched on the noisy image. The covariance parameters and the angles defining the anisotropies of the true signal and of the noise are directly used to compute the factorial kriging estimate of the true signal at each point of the grid. The output obtained from the filtering process is presented in \Cref{fig:filt_res}. As we see, the filtering process successfully extracted the true signal from the noisy observation. However we seem to obtain a smoothed version of the input: this is a consequence of the fact that the true signal is estimated through a kriging approach, which tends to yield smoothed outputs (je ne suis pas d'accord avec cette dernière phrase dans le cas du factorial kriging) \citep{wackernagel2013multivariate}.

\subsection{Real case study}

\begin{figure}[b!]
\centering
\includegraphics[width=0.8\textwidth]{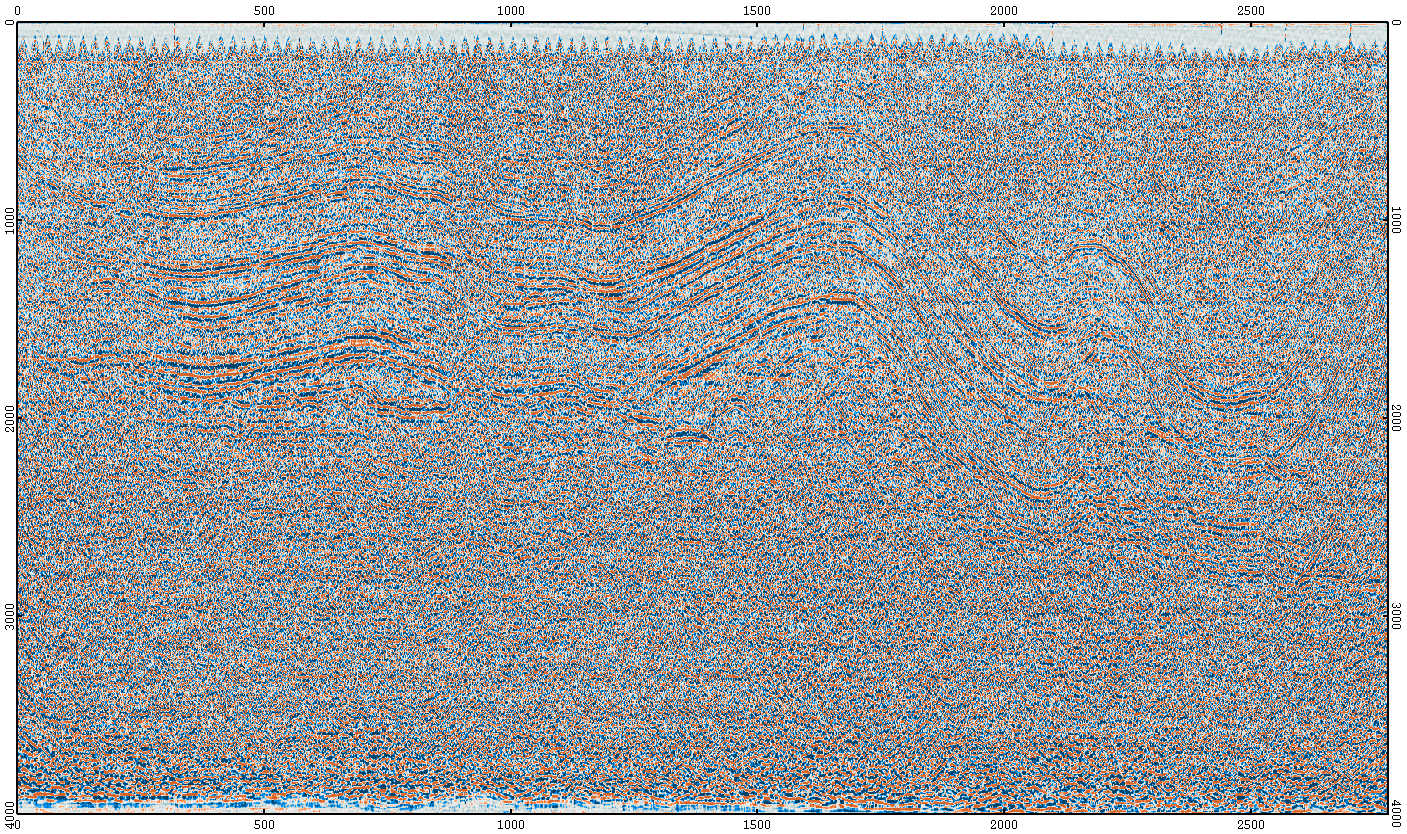}
\caption{Input seismic data from the Amadeus basin (Courtesy of CENTRAL PETROLEUM). The data form a 2778x1001 grid.}
\label{fig:filt_input_cpl}
\end{figure}

The case study corresponds to the application of the geostatistical filtering on a vintage 2D seismic line acquired in the Amadeus basin (onshore Australia), and provided by the company CENTRAL PETROLEUM. The data was originally acquired in 1966, reprocessed in 1984 and vectorized from a hard copy in 2010. It is displayed in \Cref{fig:filt_input_cpl}. As one may notice, the image is very noisy due to its long history. Moreover, in some parts of the image, the signal is almost completely attenuated by the noise: this is the case in high dip areas, where the high slope of the geological interface made it hard to retrieve a satisfying level of signal from the seismic measurements.

The first step was to derive a generic variogram model composed of the signal and noise structures. Random and linear noises were identified and characterized through stationary covariance functions. This  was done following the same approach as the one  described by \citet{magneron2009m}. In particular, a smooth component corresponding to the true signal was identified, and $5$ additional noisy components characterized by global geometric anisotropies were identified. This work was done by expert geophysicists, who used their prior knowledge of the dataset to separate what is supposed to be the noise from the signal (during the variogram modeling step).

\begin{figure}
\centering 
\includegraphics[width=0.8\textwidth]{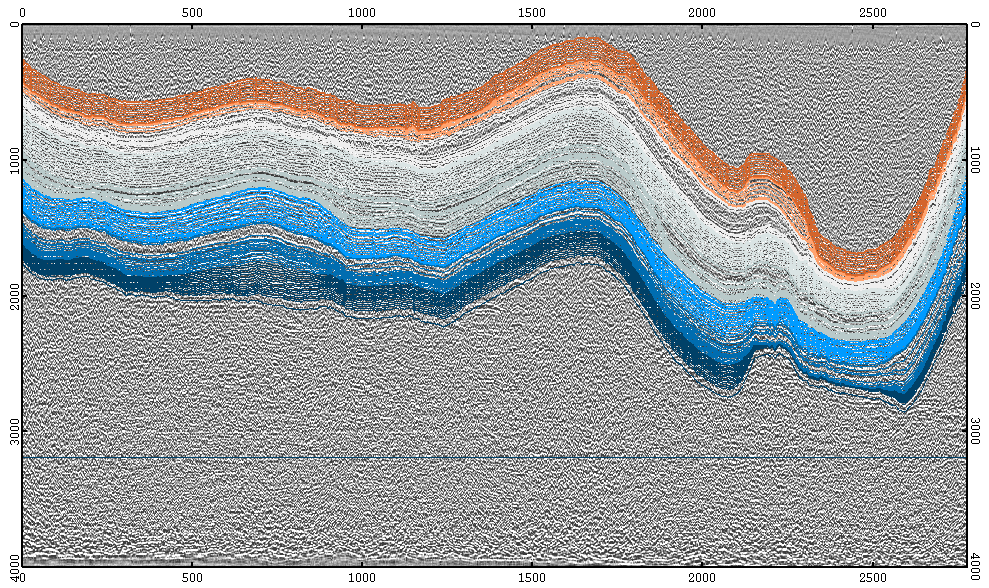}
\caption{Identification of some geological interfaces from the noisy data of the Amadeus basin using the \textit{Paleoscan}\textsuperscript{TM} software.}
\label{fig:paleo}    
\end{figure}

Then, local dips were assigned to the signal to be consistent with the geological structure. This was done using the \textit{Paleoscan}\textsuperscript{TM} software from the company ELIIS, which allows to identify the global shape of some geological interfaces from noisy seismic images (cf. \Cref{fig:paleo}). The angles describing locally these interfaces were extracted and interpolated on the whole domain. They serve as local anisotropy angles defining the signal to extract with the filtering process.

\begin{figure}
     \centering
         \begin{subfigure}[t]{0.8\textwidth}
    	\centering
        \includegraphics[width=\textwidth]{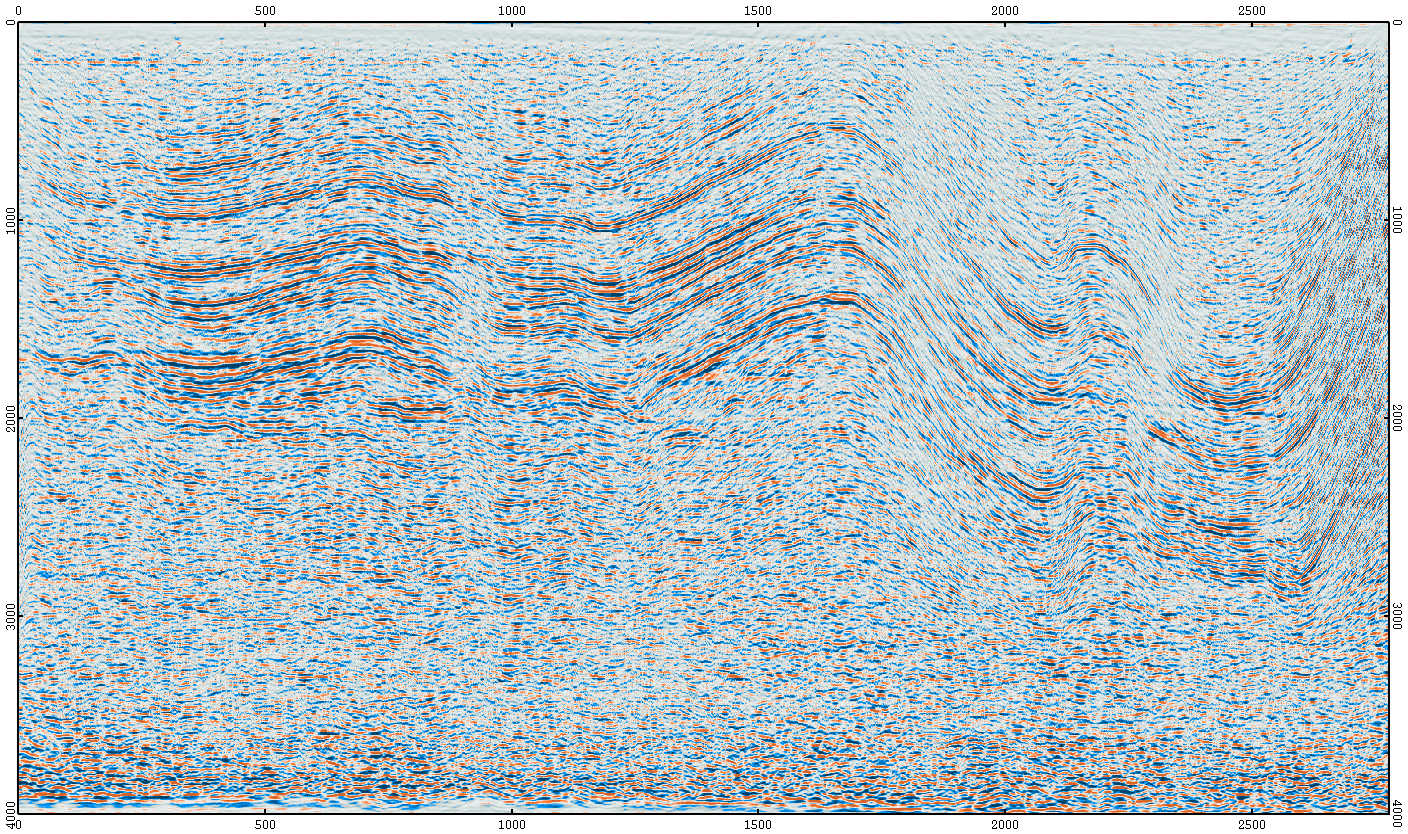}
        \caption{Filtered output.}
    \end{subfigure}\\
    \begin{subfigure}[t]{0.8\textwidth}
    	\centering
        \includegraphics[width=\textwidth]{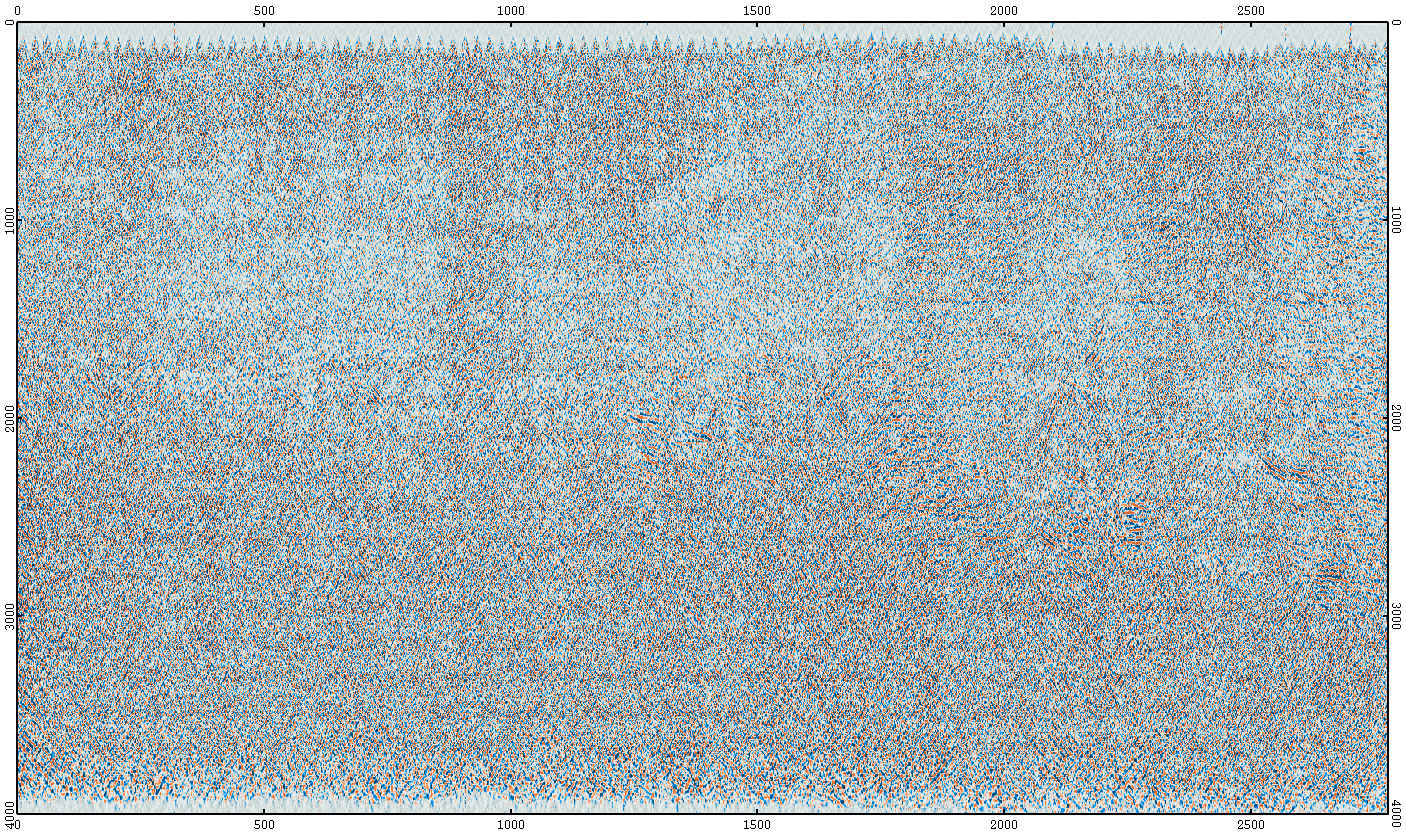}
        \caption{Noise extracted.}
    \end{subfigure}
        \caption{Results obtained from the filtering process to the noisy data of the Amadeus basin.}
    \label{fig:res_filt_cpl}
\end{figure}

Thus, with local anisotropies defined, the geostatistical filtering approach allowed the filtering out of noise while preserving the true signal. This is clearly demonstrated in \Cref{fig:res_filt_cpl}  where  there is no obvious signal remaining in the noise image. More impressive was the restoration of the signal in the high dip areas, which was only possible using our filtering approach, since the linear noise interferes strongly with the signal in these parts of the image.

The results were validated by CENTRAL PETROLEUM, and it was proposed to use this approach of geostatistical filtering solution as a valid alternative to expensive full seismic reprocessing of seismic lines. Note that in this case, the number of observation locations, which corresponds to the number of grid points, amounted to more than $2.7$ millions of points. At this level, the matrix-free approach we propose reveals its benefits: it allows us to \q{virtually} work with the huge full matrices that come with the complex non-stationary models we chose to model the complexity of the dataset.


\section{Conclusion}

In this work, we introduced a matrix-free approach to geostatistical filtering. In this approach, the noisy signal is modeled as a sum of independent Gaussian random fields defined through an expansion in a basis of eigenvalues of the Laplace (or Laplace--Beltrami) operator, defined over the spatial domain. These random fields are then numerically approximated using a finite element method. From a modeling perspective, this approach allows to easily account for non-stationary fields characterized by local anisotropies, without having to resort to neighborhood approaches or to restrict the choice of covariance model.  From a practical point of view, the recourse to finite elements allows to reduce the memory and computational needs: indeed, only a few sparse matrices need to be built and stored to run the filtering process thanks to a polynomial trick based on Chebyshev approximation. 

This approach can be extended to more than filtering, and actually paves the way to matrix-free approach to Geostatistics. Indeed, the expression of (non-stationary) covariance matrices as a matrix function depending on a sparse matrix, that we leveraged for filtering, can also be exploited to perform any other task involving covariance matrices.  For instance, we can cite the inference of the parameters characterizing a Gaussian field using a likelihood-based approach, simulations of Gaussian fields on a set of locations of a domain, and the estimation of a Gaussian field from its partial observation, using a kriging approach. Matrix-free algorithms for these purposes were actually derived by \citet{pereiraPhd2019}.


\medskip

\paragraph{Acknowledgements}
The authors would like to thank Central Petroleum for the authorization to publish the results related to the 1966 seismic line (Amadeus basin). They would also like to thank Estimages and the Chalmers AI Research Center (CHAIR) for their financial and material support.

\bibliographystyle{apalike}       
\bibliography{bib}

\end{document}